\documentclass[final,5p,times,twocolumn]{elsarticle}

\usepackage{times,amsmath,epsfig}
\usepackage{subfigure}
\usepackage{cite}
\usepackage{comment}
\usepackage{amssymb}
\usepackage{graphicx}
\usepackage{url}
\usepackage{balance}
\usepackage{amsmath} 
\usepackage{color}
\usepackage{multirow}
\usepackage{afterpage}
\usepackage{algorithm}
\usepackage[table,xcdraw]{xcolor}

\newfloat{program}{thp}{lop} 
\floatname{program}{Program}
\newcommand{\ie}{i.e.} 
\newcommand{\eg}{e.g.} 
\newcommand{\et}{et al. }

\newcommand{\new}[1]{\textcolor{black}{#1}}
\journal{Energy Journal}

\begin{document}

\begin{frontmatter}



\title{A Hybrid ICT-Solution for Smart Meter Data Analytics}

 %\author{Name\corref{cor1}\fnref{label2}}
 %\ead{email address}
 %\fntext[label2]{}
 %\cortext[cor1]{}
 %\address{Address\fnref{label3}}
 %\fntext[label3]{}



\author{Xiufeng Liu}
\cortext[cor1]{Corresponding author}
\ead{xiuli@dtu.dk}

\author{Per Sieverts Nielsen}
%\ead{pernn@dtu.dk}

\address{Technical University of Denmark}

\begin{abstract}
Smart meters are increasingly used worldwide. Smart meters are the advanced meters capable of measuring energy consumption at a fine-grained time interval, \eg, every 15 minutes. Smart meter data are typically bundled with social economic data in analytics, such as meter geographic locations, weather conditions and user information, which makes the data sets very sizable and the analytics complex. Data mining and emerging cloud computing technologies make collecting, processing, and analyzing the so-called {\em big data} possible. This paper proposes an innovative ICT-solution to streamline smart meter data analytics. 
The proposed solution offers an information integration pipeline for ingesting data from smart meters, a scalable platform for processing and mining big data sets, and a web portal for visualizing analytics results. The implemented system has a hybrid architecture of using  {\em Spark} or {\em Hive} for big data processing, and using the machine learning toolkit, {\em MADlib}, for doing in-database data analytics in {\em PostgreSQL} database. This paper evaluates the key technologies of the proposed ICT-solution, and the results show the effectiveness and efficiency of using the system for both batch and online analytics.
\end{abstract}


\end{frontmatter}

\section{Introduction}
\label{sec:introduction}

Today smart meters are increasingly used worldwide for their ability to provide fine-grained automated readings without visiting customer premises. Smart meters collect energy consumption data at a regular time interval, usually every 15 minutes or hourly. An ICT solution provides the platform for analyzing the collected meter readings, which has become an indispensable tool for utilities to run smart grids. Smart meter data analytics can help utilities better understand customer consumption patterns, provision energy supply to the peak demand, detect energy theft, and provide personalized feedback to customers. Moreover, the government can take decisions for the future grid development based on analytics results. For customers, smart meter data analytics can help them better understand their energy consumption, save energy, and reduce their bills. Data analytics thus is seen so important in smart grid management that the global market has been growing rapidly in recent years, and the market is expected to reach over four billion dollars annually in 2020 \citep{msw10}.  
 
Various algorithms have been proposed for analyzing smart meter data, mainly in the smart grid literature, including those for electricity consumption prediction, consumption profile extraction, clustering of similar customers, and personalized customer feedback on energy savings. Nevertheless, smart meter analytics applications have been underdeveloped until recently when some database vendors started to offer smart meter analytics software, \eg, SAP and Oracle/Data Raker. And, several startups are seen investing in this area, e.g., C3Energy.com and OPower.com. Furthermore, some utilities, such as California's PG\&E4, are now providing on-line portals where customers can view their electricity consumption and compare with their neighborhood's average. However, these systems and tools focus on simple aggregation and ways of visualizing consumption. The details of how they are implemented are not disclosed. A lot of work is still required to build practical and scalable analytics systems for handling smart meter data, characterized by big volume and high velocity. 
 
In this paper, we propose a hybrid ICT-solution for streamlining smart meter data analytics extended from our conference paper \citep{sdewe2015}. This solution is built by compiling our recent scalable data processing framework, {\em BigETL} \citep{bigetl}, and our prototype system, {\em SMAS} \citep{smas2015}. The proposed ICT-solution aims at facilitating the whole process of smart meter data analytics, including data ingestion, data transformation, loading, analyzing, and visualization. Utilities and customers can get near real-time information through these stages. The ICT-solution has a hybrid system architecture, consisting of the building blocks Spark and Hive in the data processing layer, and PostgreSQL with MADlib \citep{madlib} in the analytics layer. The design considers the support for high-performance analytics queries, \ie, through RDBMS, and the support for big data analytics, \ie, through Spark and Hive. We decouple data ingestion, processing, and analytics into three different layers to ease the use and further development. Meter data go through the three layers from the sources to the visualization on a web portal. The processing layer is an open platform that can integrate various user-defined processing units, such as the modules of data transformation, data anonymization, and anomaly data detection. The analytics layer currently supports a variety of functionalities and analytics algorithms, including viewing time-series at different temporal aggregations (e.g., hourly, daily, or weekly), load disaggregation, consumption pattern discovery, segmentation, forecasting and customer feedback. It is also easy to extend the analytic layer by adding new algorithms. 
 
In short, we make the following contributions in this paper: 1) we propose a hybrid ICT-solution of making the advantages of different technologies for streamline meter data analytics process; 2) we implement a data processing platform that can be easily extended by adding new processing units and algorithms; 3) we implement an analytics platform that can support both supply- and demand-side analytics. It can help utilities manage energy and help customers save money; and 4) we evaluate the technologies that constitute the proposed ICT platform, and discuss which technology to be used and when.

The remaining part of this paper is structured as follows. Section~\ref{sec:relatedwork} summarizes the related work; Section~\ref{sec:solution} details the proposed ICT-solution; Section~\ref{sec:evaluate} evaluates the technologies which constitute the ICT-solution;  Section~\ref{sec:conclusionandfuturework} concludes the paper with the directions for future work.

\begin{figure*}[htp]
%\vspace{-5pt}
\centering
\includegraphics[width=0.7\textwidth]{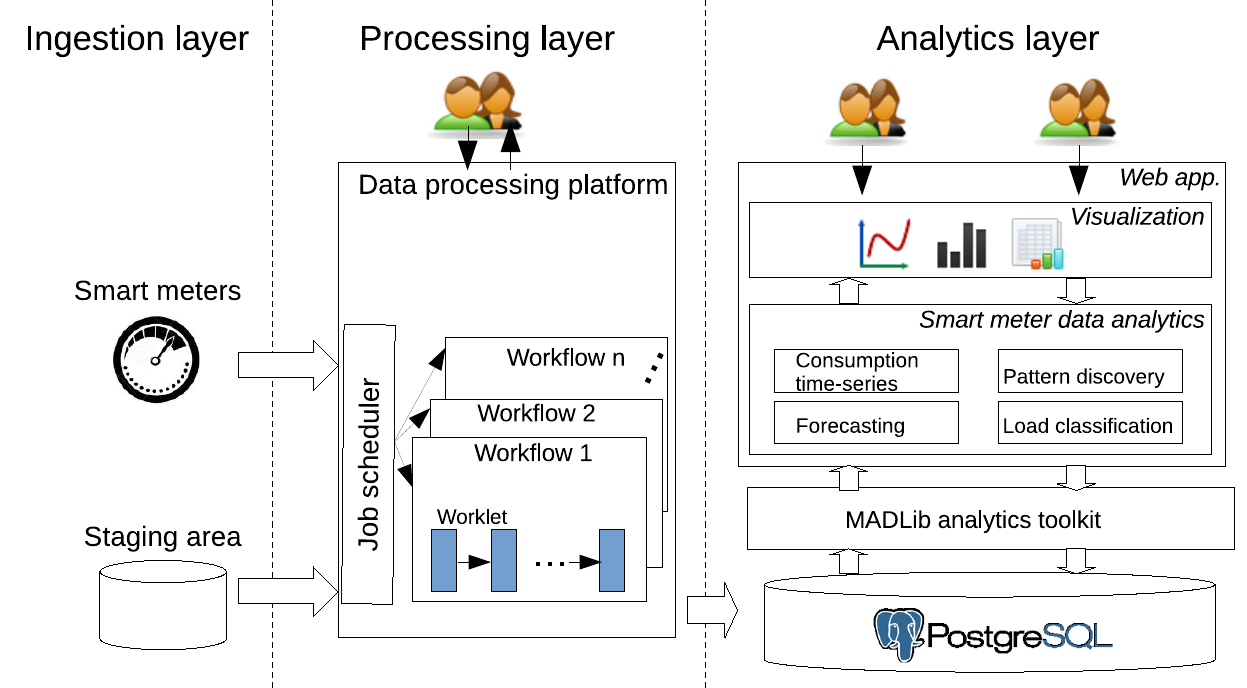}
%\vspace{-5pt}
\caption{The system architecture of smart meter data analytics system (the analytic layer is reproduced from  \citep{smas2015})}
\label{fig:systemarchitecture}
%\vspace{-5pt}
\end{figure*}

\section{Related Work}
\label{sec:relatedwork}
{\em Systems and Platforms for Smart Meter Data Analytics.} 
The traditional technologies, such as R (S-PLUS), Matlab, SAS, SPSS and Stata,  can be used in smart meter data analytics to support numeric computing and comprehensive statistical analysis. In-memory, in-database, and parallelism are the recent trend of analytics technology development. According to our benchmarking \citep{benchmark2015}, main-memory based systems, such as KDB+ \citep{kdb} and SAP HANA \citep{farber2012}, and in-database machine learning toolkit, such as MADlib \citep{madlib}, are good options for smart meter analytics. The two distinct distributed computing platforms,  Hive and Spark (built on top of Hadoop),  are suitable for big data analytics. In this paper, we implement the ICT-system with the hybrid technologies, including Hive, Spark and PostgreSQL/MADlib, which enable us to analyze data in a database, in memory or in a cluster. 

The systems or prototypes for smart meter data analytics emerge in both industry and academia. The companies we mentioned in Section~\ref{sec:introduction} developed intellectual property products, but the implementations of the systems and analytics algorithms used are not open to the public.
\new{Personal \et implemented a tool to assess smart grid goal based on their developed key performance indicators \citep{Personal2014}}. Nezhad \et developed an open source smart meter dashboard in their research work, called {\em SmartD} \citep{smartd}, which is orthogonal to our work of the analytics layer. But, our system provides more comprehensive functionalities, and we provide a complete solution for smart meter data analytics, including data ingestion, transformation, analyzing and visualization. Liu \et use analytics procedures in Hive to process smart grid data, and use an RDBMS to cope with daily data-management transactions on the information of meter devices, users, organizations, etc. \citep{liu2014}. This is somewhat similar to our architecture, but our main focus is to streamline the whole process and to cater for different user requirements by using hybrid technologies. Furthermore, our platform is designed to be easily integrated with different data processing units and algorithms. Besides, the work primarily studies how to efficiently retrieve smart meter data from Hive, but it focuses on simple operational queries rather than in-depth analytics that we address in our system. Beyond the electricity sector, smart meter analytics systems and applications are also developed in the water sector, e.g., {\em WBKMS} \citep{stewart2015}, a web-based application for providing real-time information of water consumption; and {\em Autoflow} \citep{khoi2015}, a tool for categorizing residential water consumption. Water data analytics is one of our planned features. The two existing works provide a good reference for us to design and develop the architecture and algorithms in the future.

{\em Benchmarking Smart Meter Data Analytics.} 
Arlitt \et implement the toolkit, {\em IoTAbe\-nch}, for benchmarking the analytics algorithms of  Internet of Thing (IoT) \citep{arlitt2015}. They evaluate six analytics algorithms on an HP Vertica cluster platform using synthetic electricity data. Keogh \et discuss benchmarking time-series data mining, and evaluate different implementations of time series similarity search, clustering, classification and segmentation \citep{keogh2003}. Anil benchmarks data mining operations for power system analysis \citep{anil2013}, which analyzes voltage measurements from power transmission lines. These works, however, only focus on benchmarking the analytics algorithms, but not the systems in the underlying. Our previous work \citep{benchmark2015} benchmarks four representative algorithms of smart meter data analytics, and five categorized technologies, Matlab, KDB+, PostgreSQL/MADlib, Spark and Hive. They represent the technologies of traditional (Matlab), in-memory (KDB+ and Spark), in-database (PostgreSQL/MADlib), in-memory distributed (Spark) and Hadoop-based (Hive). The benchmarking results are the foundation for implementing this system, \ie, provide the guideline for choosing the appropriate technology for an analytics requirement. 
 
{\em Smart Meter Data Analytics Algorithms.} Two broad categories of smart meter data analytics applications are widely studied, which are consumer- and producer-oriented. Consumer-oriented applications aim to provide feedback to end-users on reducing energy consumption and saving money,  \eg, \citep{msw10,birt,rvn10}. Producer-oriented applications are for utilities, system operators and governments, which provide information of consumers, such as their daily habits for the purposes of load forecasting and clustering/segmentation, \eg, \citep{msw10,smartd,zeng20141,acf12,Gebru,ar13,omid,beckel2014,chicco,espinoza,frv05,ghe11}. From a technical standpoint, both of the above classes of applications perform two types of operations: extracting representative features, \eg, \citep{birt,omid,espinoza,frv05}, and finding similar consumers based on the extracted features, \eg, \citep{rvn10,acf12,chicco,smith,tsekouras}. Household electricity consumption can be broadly decomposed into the temperature-sensitive component, \ie, the heating and cooling load, and the temperature-insensitive component (other appliances). Thus, the representative features include those which measure the effect of outdoor temperature on consumption \citep{birt,rvn10,albert_bigdata} and those which identify consumers' daily habits regardless of temperature \citep{acf12,omid,espinoza}, as well as those which measure the overall variability, \eg, consumption histograms \citep{Gebru}. Some of the above existing algorithms have been integrated into our system, as well as new algorithms implemented by us, which are used to study the variability of consumption, load profiling, load segmentation, pattern discovery, load disaggregation, and load similarity.    
 
\section{\new{Methodology}}
\label{sec:solution}
\subsection{Solution Overview}
\label{sec:systemoverview}
Figure~\ref{fig:systemarchitecture} shows the system architecture of the proposed ICT-solution. The system is implemented by integrating our two sub-systems, {\em BigETL} \citep{bigetl} and {\em SMAS} \citep{smas2015}, which will be described in the next two sections. The system consists of three layers including data ingestion, data processing, and data analytics, each of which represents a separate functional system for meeting the overall requirement of streamlining the whole process. The leftmost is the data ingestion layer extracting real-time data from smart meters directly, or bulk data from a staging area. This layer connects data sources to the subsequent processing layer using data extraction programs. 

The middle layer is responsible for pre-processing data, such as transformation and cleansing. This is done through the so-called {\em workflow}, which is composed of several chained processing units, {\em worklets}.  A worklet is run in a different processing system in the underlying (we will detail it shortly). A workflow is scheduled to run once or repeatedly at a specified time interval, such as minutely, hourly, daily, weekly or monthly. This layer can also manage multiple workflows running simultaneously. The worklets in a workflow are executed in a sequential order, each of which is responsible for a particular task. For example, a batch processing workflow may consist of a worklet for extracting data from a source system and writing to staging area; a worklet for cleansing the data and writing the cleansed data into the data warehouse;  a worklet for housekeeping the staging area; and a worklet for sending messages when the workflow job is end. 

The rightmost is the analytics layer, which is a web application composed of an application server (Tomcat), a visualization engine (Highcharts), various analytics algorithms (implemented using the open source in-database analytics library, MADlib \citep{madlib}), and a data warehouse (PostgreSQL). This layer has a web-based interface for users doing interactive analytics.

\subsection{BigETL: The System for the Pre-processing layer}
\label{sec:processinglayer}
\begin{figure}[htp]
%\vspace{-5pt}
\centering
\includegraphics[width=0.42\textwidth]{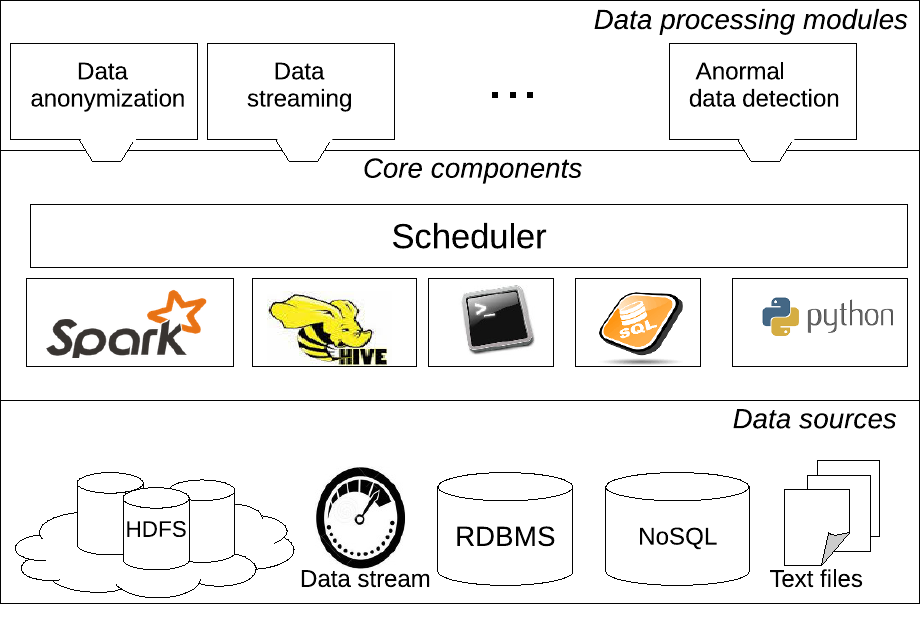}
%\vspace{-5pt}
\caption{The building blocks of BigETL}
\label{fig:dataprocessinglayer}
%\vspace{-20pt}
\end{figure}

The ICT-solution uses {\em BigETL} in its data processing layer. BigETL is developed based our previous works \citep{etlmr2011,tldks2011,pvldb2011,cloudetl2014} for handling scalable and streaming data (the source code is available at \url{http://github.org/xiufengliu/BigETL}). Figure~\ref{fig:dataprocessinglayer} shows the building blocks, including the interfaces for supporting various data sources, data processing systems (incl. Spark, Hive, Linux Shell, Java Virtual Machine, and Python), a job scheduler, transformation and online analytics units. A processing unit is a program implemented for a specific purpose, such as data cleansing, data transformation, data anonymization, or streaming data mining (e.g., anomaly detection). The unit is scheduled to run on an underlying data processing system. BigETL provides the interface for integrating a data processing unit. It also supports reading data from various data sources and writing data into different data management systems, which is simply through implementing the corresponding reading and writing application programming interfaces.

\subsubsection{Data Stream Processing}
\begin{figure}[htp]
%\vspace{-15pt}
\centering
\includegraphics[width=0.49\textwidth]{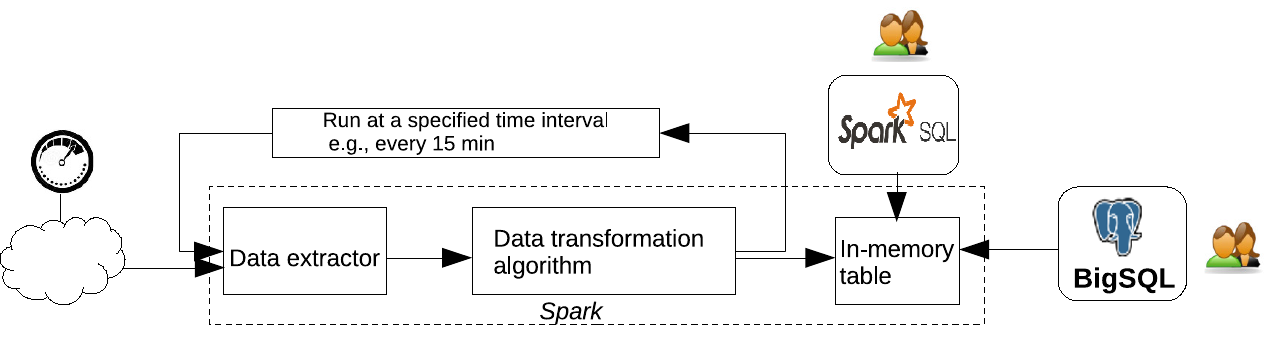}
%\vspace{-5pt}
\caption{Near real-time data stream processing}
\label{fig:realtimeprocessing}
%\vspace{-15pt}
\end{figure}

We use Spark Streaming \citep{sparkstreaming} (a built-in component of Spark) to process data stream (see Figure~\ref{fig:realtimeprocessing}), e.g., the readings from smart meters. Meter readings can be extracted periodically, \eg, typically every 15 minutes or one hour. BigETL, however, can extract data from any sources as long as the corresponding data extractors are implemented. When the data are extracted and read into Spark streaming, the data are created as Discretized Stream (DStream), a continuous stream of data, received from a data source, or a processed data stream generated by transformation operators.  A DStream is represented by a continuous series of resilient distributed data objects (RDDs) in Spark, which is an abstraction of an immutable distributed data set. Therefore, DStreams are the fault-tolerant collections of objects partitioned across cluster nodes that can be acted on in parallel. A number of operations, called {\em transformations}, can be applied to a DStream, including map, filter,  groupBy, and reduce, etc., as well as windowing operations with a user-specified size and slide interval. When the data have been processed, the cleaned and well-formatted data are temporarily kept in an in-memory table in Spark, which is also an RDD object but given the names to its attributes to improve interoperability through SQL. Users can do ad-hoc queries by SQL statements issued on a web-based user interface. The query results are shown in table or chart format on the web portal. Through the interactive queries, users can check the results instantly, verify and improve their query statements. In the end, the in-memory data are persisted to the PostgreSQL data warehouse for end-user interactive analytics. But, if an end user wants to view the freshest data in Spark, they still can query from the analytics layer. In the underlying, the queries are sent to Spark through a middleware, BigSQL \citep{bigsql}, which bridges PostgreSQL and Spark.  

\subsubsection{Batch Processing}
BigETL supports Hive as the batch processing system to deal with scalable data sets. Hive is an open-source SQL-based distributed data warehouse system built on top of Hadoop MapReduce framework \citep{hive2010}, which now is widely used in big data analytics. Hive supports the SQL-like script language, HiveQL, to query data stored in a cluster. Internally, Hive translates an SQL statement into Hadoop MapReduce jobs by an SQL-to-MapReduce translator and runs in parallel in the cluster. This greatly lowers the barrier to using Hadoop, thus, a user can use Hadoop as long as (s)he is familiar with SQL script language. Due to the low latency of Hadoop, and the append-only Hadoop distributed file system (HDFS), Hive is only suitable for the situation where large-scale data are analyzed, fast response time is not required, and no frequent data updates are needed. Therefore, we choose Hive as the off-line analytics system for big data. Analytics queries are run as MapReduce jobs, and the results are written into a Hive table, a logical data organization structure on HDFS. The results can be exported into PostgreSQL for interactive online analytics queries and visualization.

\subsubsection{Job Scheduling}
The data processing layer supports different workflows running on the same platform scheduled by a scheduler. To coordinate the jobs, and control the use of computing resources (e.g., memory and CPU), BigETL uses a centralized job scheduling system to schedule the runnings of all workflows. The system adopts two types of scheduling methods, \ie, {\em deterministic} and {\em un-deterministic}. The deterministic method is used to schedule a workflow to run exactly at the time specified by users, \ie, the starting time of a workflow is deterministic and remains the same for repeating executions. The workflows scheduled by this method are typically those that run on the processing systems in a single server environment. The un-deterministic method, on the other hand, is used to schedule the workflows running in a cluster environment, \ie, the actual starting time of a job is not necessary at the time specified by users, but usually later than the specified time. The scheduling method ensures that only one job is running in a cluster at any point of the time. The reason is that if multiple jobs are submitted to the same cluster, the submitted jobs will compete for the limited computing resources, and the cluster may become unstable, \eg, some runtime exceptions might be thrown, such as the notorious {\em out of memory} exception. In this method, a queue is used to accommodate all submitted jobs, and the jobs run according to first-in-first-out (FIFO) strategy. Although Spark and Hadoop have their own job schedulers, we implement this scheduling system for better controlling the runnings of workflows, \eg, we can chain multiple workflows and run in a sequential order in our platform.   

\subsection{SMAS: The System for the Analytics Layer}
\label{sec:analyticslayer}

The ICT platform uses SMAS in its analytics layer, which was developed in our previous work \citep{smas2015} for analyzing smart meter data (the source code is available at \url{http://github.org/xiufengliu/smas}). As shown in Figure~\ref{fig:systemarchitecture}, this layer is a web application for users to do interactive analytics using the smart meter data in the PostgreSQL data warehouse with a variety of analytics algorithms implemented using MADlib.
\begin{table*}[htp]
%\vspace{-15pt}
\centering
\caption{Table layout for storing time-series data, {\em tbl\_hourlyreading}}
\begin{tabular}{p{2.1cm}p{2.1cm}p{2.1cm}p{3cm}p{3.3cm}}
\hline
MeterID & Temperature & Reading & Readtime & TempIndependentLoad \\ \hline
100  & 5.7 & 2.3 & 2014-01-01 00:00:00 & \\
100  & 5.6 & 1.8 & 2014-01-01 01:00:00 & \\
100  & 5.6 & 2.1 & 2014-01-01 02:00:00 & \\
...  & ...  & ... & ...                &  \\  \hline
\end{tabular}
\label{tab:tablelayout}
%\vspace{-10pt}
\end{table*}
\begin{figure*}[htp]
%\vspace{-20pt}
\centering
\includegraphics[width=0.8\textwidth]{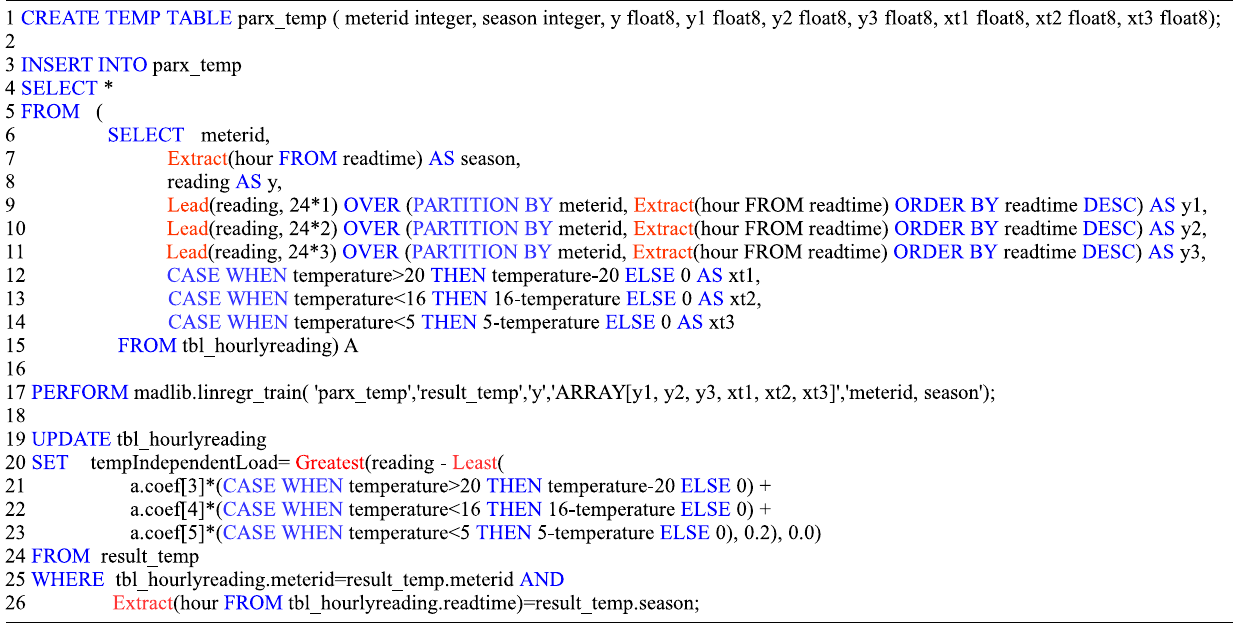}
\caption{Extracting temperature-independent consumption using Pl/PgSQL and MADlib}
\label{fig:usingmadlib}
%\vspace{-10pt}
\end{figure*}

\subsubsection{In-database Analytics using MADlib}  
\label{sec:indbanalysis}
In traditional execution of analytics using SAS, R, Matlab and Proc-SQL there is significant data movement from a database to the analytics tools. The part of the workload related to data movement is often quite substantial; thus, a new trend is to push analytics algorithms into a database. In SMAS, all the algorithms are implemented as stored procedures using MADlib analytics library and Pl/PgSQL script programming language and do the {\em in-database} analytics in PostgreSQL. The used analytics functions from MADlib include linear regression, k-means, histogram, and ARIMA. MADlib also offers many other analytics functions, including logistic and multinomial logistic regression; elastic-net regularization for linear and logistic regressions; association rules; cross-validation; matrix factorization methods; LDA, SVD and PCA; and more \citep{madlib}. All the MADlib functions are used through a pure SQL interface.

We now show an example on how to implement an algorithm using MADlib and Pl/PgSQL. This algorithm is called periodic auto-regression with eXogenous variables (PARX) \citep{omid}, used to extract the load regardless of any exogenous variables affected. We take residential electricity consumption as the example. The consumption may be affected by the exogenous variables, such as family size, house area, outdoor weather temperature, and so on. For simplicity, this example only considers the impact of weather temperature, \ie, given an hourly electricity consumption time series of a customer over time. We use this algorithm to determine, for each hour, how much load is temperature-independent. Due to the habit of a customer, the daily load may follow a certain pattern, \eg, having a morning peak at 7-8 o'clock after getting up, and evening peak at 17-20 o'clock for cooking after work. This algorithm takes 24 hours as a period, and the hours of a day at $t$, \ie, $t=0...23$, as the seasons, and uses the previous $p$ days at the hour $t$ for auto-regression. Thus, the model at the $s$-th season and at the $n$-th period is formalized as:   

\begin{equation}
    Y_{s,n} = \sum_{i=1}^{p} \alpha_{s,i} Y_{s,n-i}  + \beta_{s,1} XT1 + \beta_{s,2} XT2 + \beta_{s,3} XT3 + \epsilon_s, \enspace s \in t
    \label{eq:parx}
\end{equation}
where $Y$ is the data point in the time-series; $p$ is the number of order in the auto-regression; $XT1, XT2$ and $XT3$ are the exogenous variables accounting for temperature, defined in the equations of $(2) - (4)$; $\alpha$ and $\beta$ are the coefficients; and $\epsilon$ is white noise.

\begin{equation}
	XT1 = \begin{cases}
T-20 & \text{ if } T > 20 \\ 
0 & \text{ otherwise }  
\end{cases}
\end{equation}
\begin{equation}
	XT2 = \begin{cases}
16-T & \text{ if } T < 16 \\ 
0 & \text{ otherwise }  
\end{cases}
\end{equation}

\begin{equation}
	XT3 = \begin{cases}
5-T & \text{ if } T < 5 \\ 
0 & \text{ otherwise }  
\end{cases}
\end{equation}
The variables represent the cooling (temperature above 20 degrees), heating (temperature below 16 degrees), and overheating (temperature below 5 degrees) slopes, respectively.

Table~\ref{tab:tablelayout} is the layout for storing hourly time series in PostgreSQL database. This table contains the time series of residential electricity consumption, as well as hourly temperature time series obtained from a local weather station. Initially, we leave the temperature-independent load empty (see the column \texttt{TempIndependentLoad}), which will be updated later by the algorithm (see Figure~\ref{fig:usingmadlib}). The implementation is shown in Figure \ref{fig:usingmadlib} (only the body of the stored procedure is listed). Since MADlib does not provide the built-in PARX implementation, we, instead, implement the algorithm using the provided multiple linear regression function, which takes the values of $Y$, $XT1, XT2$ and $XT3$ as the inputs of the variables. Lines~1--15 show the process of preparing the variable values for the algorithm input in a temporary table, where we set the order $p=3$ for the auto-regression part. Line 17 executes MADlib regression function where the output is saved to \texttt{result\_temp} table. Lines~20--26 compute the temperature-independent load according to the equation \ref{eq:parx}, and updates to the time-series table.

We have evaluated the validity of the coefficients of PARX model in \ref{apx:parx}. In our platform, the PARX algorithm is used for hourly-ahead short-term forecasting of customer energy demand. Also, we use it in consumption disaggregation. Since this algorithm considers the impact of weather temperature, the consumption can be disaggregated into the temperature-independent part, and the temperature-dependent part, respectively. The temperature-independent part is the activity load caused by the people's indoor activities, such as cooking, lighting, and laundry. The activity load patterns are clustered in our system, which allows utilities to discover the customers with similar living habits to provide personalized energy-related recommendations.

\subsubsection{System Functionalities}
The system supports both supply-side and demand-side analytics. Through the supply-side analytics, utilities can optimize smart grid, plan for the future, and provision for the peak of demand. Through the demand-side analytics, utilities can better understand customer consumption, and provide personalized services to customers. Customers can better understand their own consumption as well, which help them save energy. Followings are the functionalities implemented so far.

 {\em Consumption Analysis.} Utilities can view the consumption at different granularities with respect to time and geographic location dimensions, and view the aggregated consumption with respect to the functions, such as sum, average, min or max.  Utilities can also compare the consumptions of any two individual customers or feed areas. 

 {\em Consumption Pattern Discovery.} 
Smart meter time-series data reflect the load influenced by various factors, including indoor consumer activity, outdoor weather temperature, and the appliances used. Consumption pattern discovery helps utilities  better understand consumption practices of customers so that they can provide customers appropriate recommendations for energy saving. In pattern discovery, the system provides the view of learning energy consumption distribution of a customer, the view of daily load shapes according to weekdays and weekends/holidays, and the view of disaggregated consumption in terms of base load and activity load (see the bottom chart in Figure~\ref{fig:patterndiscovery}).

 {\em Segmentation.} For utilities, an important task is segmenting customers according to the consumption and load profiles to carry out precise marketing communication, e.g., promote the most appropriate energy-saving program to a targeted segmentation. The system can cluster customers according to the extracted consumption features, including base load, activity load, heating and cooling gradients; and the daily load profiles; and average daily/weekly/monthly load shapes. The system can display clustered customers on Google map, indicating by a different color for each cluster (see Figure~\ref{fig:segmentation}).

{\em Forecasting.} The energy industry is reliant on balancing energy supply and demand and is thus required to predict energy consumption. For instance, by predicting the periods of peak demand, utilities can avoid distribution failure by upgrading the infrastructure for more capacity, using dynamic pricing to incentivize customers to lower energy usage during peak times, or giving the recommendation of shifting from the peaks. The system provides the forecasting based on an individual customer, a feed area,  all customers and feed areas; and the time granularity of daily, weekly, and monthly. The supported forecasting algorithms include PARX, ARIMA, and exponential smoothing HoltWinters.

{\em Online anomaly detection.} Customers increasingly demand to be able to  monitor their energy consumption in near real time. Online anomaly detection allows customers to detect the unusual consumption compared with their  consumption history or their neighborhood. Customers can get the notifications for anomalous consumption.  

{\em Feedback Service.} Feedback service allows utilities to set the rules of sending alerting messages to customers. Utilities can provide a comparative feedback via ranking: what is a customer's rank within the neighborhood and the whole city in terms of overall consumption, base load, heating gradient and cooling gradient, and so on.  Once a feedback rule is met, the system will automatically send messages with a pre-set time interval.

{\em View my consumption and compare with neighborhoods.} Customers can explore their consumption data at different granularities of the time dimension, \ie, hourly, daily, weekly or monthly (see Figure~\ref{fig:myconsumption}). A customer can also compare the consumption with the average of his/her neighborhoods (not allowed to compare with an individual in the neighborhood, due to the privacy). By the comparison, a customer may find the cause of high consumption, and improve energy efficiency, e.g., high consumption may be due to air conditioners not being set to a higher temperature during nights or using old inefficient air conditioners.

\subsubsection{Integrated Analytics Functions}
We list the analytics functions used in different modules and their descriptions in Table~\ref{tbl:functionlist}. Since the  analytics framework is built within PostgreSQL database, except the anomaly detection module, we use off-the-shelf functions offered by MADlib and PostgreSQL as possible, \eg, the aggregation functions of PostgreSQL such as min, max, avg, sum, etc; and the analytics functions of MADlib, such as histogram, quantile, linear regression, k-means, ARIMA etc. In addition, we implement our analytics algorithms as PostgreSQL stored procedures, including three-line and PARX algorithms. For the online anomaly detection, we implement the programs using Spark Streaming and the analytics functions from Apache common math library \citep{apachemath}. 
\begin{table*}[t]
\centering
\caption{List of integrated analytics functions in different modules}
\label{tbl:functionlist}
\begin{tabular}{p{3cm}p{13cm}}
\hline
\multicolumn{1}{c}{\textbf{Analytics module}}     & \multicolumn{1}{c}{\textbf{Functions and description}}                                                             \\ \hline
\multirow{4}{*}{Load profiling}                    &  Aggregation functions: percentile, mean, min, max, \& median                                                \\
                                                   &  Histogram: Study consumption variability                                                                    \\
                                                   &  Three-line algorithm: Study the thermal sensitivity of energy consumption                                   \\
                                                   &  Multiple regression: Used in three-line algorithm                                                    \\
\multirow{2}{\linewidth}{Pattern discovery \& \newline segmentation} &  Aggregation functions: percentile mean, min, max, \& median                                                 \\
                &  k-means: Cluster customers with a similar consumption pattern                                               \\
Load disaggregation                                & PARX: Disaggregate consumption into temperature-independent (activity load) and temperature-dependent loads \\
Forecasting                                        & PARX, ARIMA and Holt-Winters: Short-term forecasting                                                        \\
Anomaly detection                                  & Aggregation and Gaussian distribution functions                                                             \\ \hline
\end{tabular}
\end{table*}

\section{\new{Results and Discussion}}
\label{sec:evaluate}
As mentioned in Section \ref{sec:solution}, the proposed ICT-solution supports both batch and near real-time analytics for smart meter data. We now evaluate the proposed solution and discuss the results using the four illustrative examples in Figure~\ref{fig:illustrativeexamples}.

\subsection{Illustrative Examples}
\begin{figure*}[t!]
 \centering
  \subfigure[Consumption variability analytics]{
\epsfig{file=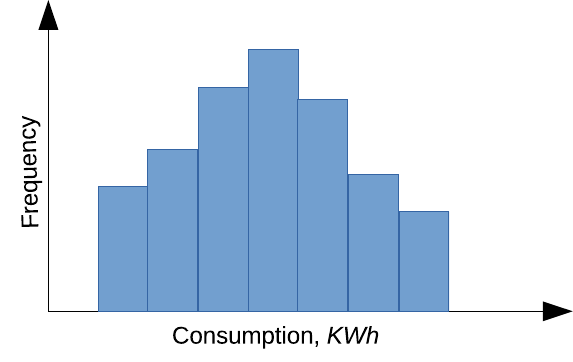, scale=1.1}
   \label{fig:histogram}
   }
   ~
 \subfigure[Thermal sensitivity analytics]{
\epsfig{file=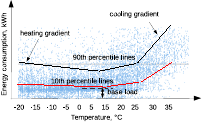, scale=3.5}
   \label{fig:3line}
   }
   
 \subfigure[Daily load profile analytics]{
  \epsfig{file=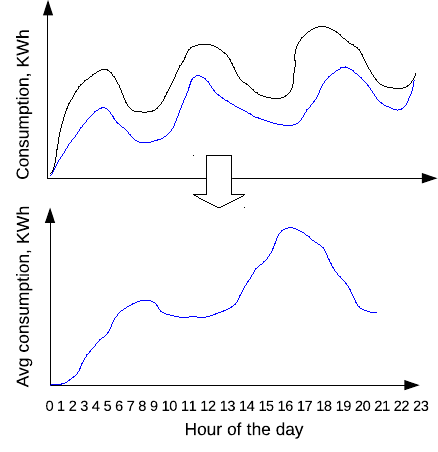, scale=1.3}
   \label{fig:parx}
   }
   ~
    \subfigure[Anomaly analytics]{
  \epsfig{file=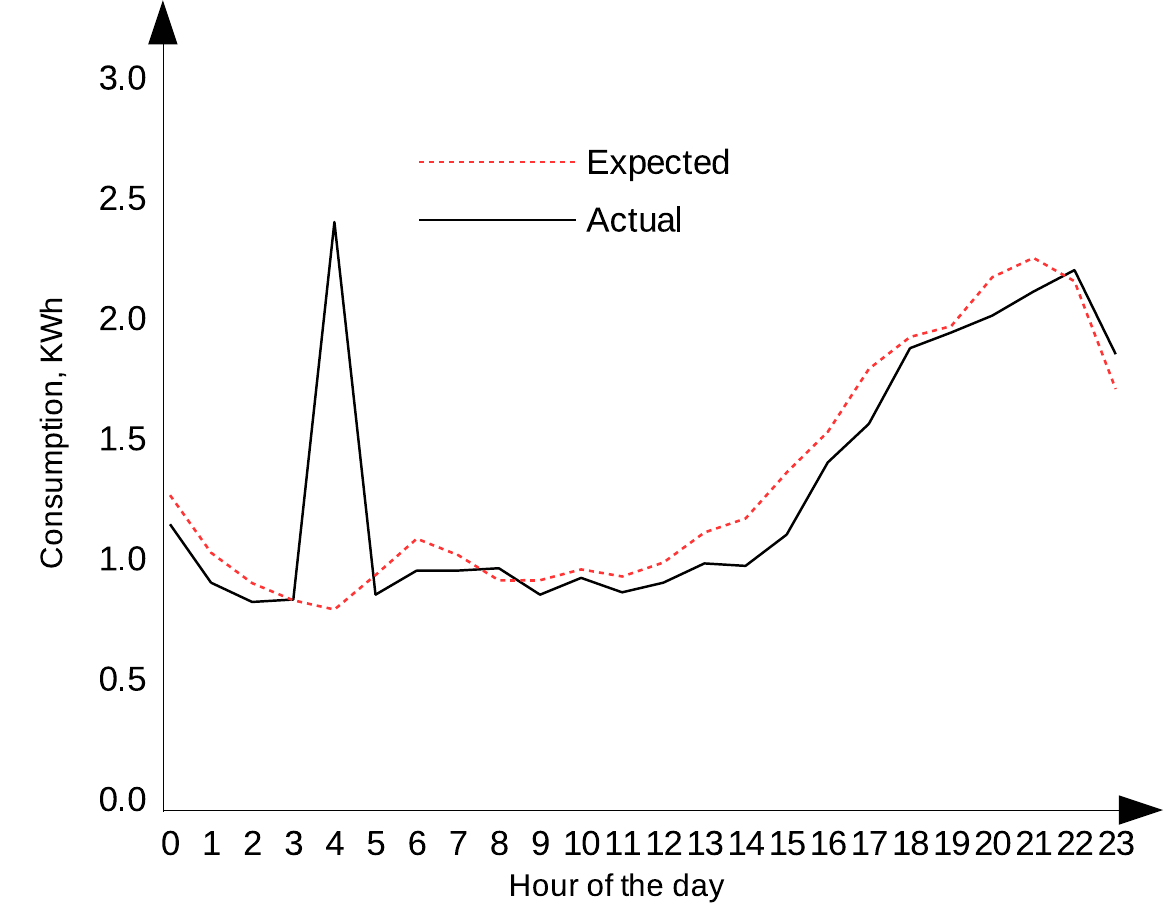, scale=0.58}
   \label{fig:anomalydetect}
   }
  \caption{Illustrative examples of smart meter data analytics (The figures are reproduced from  \citep{smas2015,benchmark2015})}
  \label{fig:illustrativeexamples}
\end{figure*}

The first example is studying {\em consumption variability} of each customer. In smart grid management, utilities must be provisioned for peak demand. Therefore, it is important for utilities to identify consumers with highly variable consumption and offer them incentives to smooth out their demand. Utilities can run histogram on the hourly consumption of each customer to learn the variability (see Figure~\ref{fig:histogram}). For simplicity, we use equi-width histograms (rather than equi-depth) in our evaluation, and we always use ten buckets.

The second example is studying {\em thermal sensitivity} of residential electricity consumption of each customer. For example, in winter and summer, consumption rises as temperatures become more extreme due to heating and cooling. Consider the scatter plot shown in Figure~\ref{fig:3line}, with temperature on the X-axis and consumption on the Y-axis. Each point on the scatter plot corresponds to a particular hourly consumption value and the corresponding temperature at that hour (for the same customer). We implement a recent algorithm \citep{birt} that computes a piecewise linear regression model to characterize thermal sensitivity. The algorithm computes two regression models: one corresponding to the 90th percentile consumption at each temperature, and the other corresponding to the 10th percentile at each temperature. The models reveal several interesting features for each customer. For example, the slope of the 90th percentile line corresponding to high temperature is the cooling gradient, and the slope of the line corresponding to low temperature is the heating gradient. Furthermore, the height of the 10th percentile lines at the lowest point is the base load, which corresponds to the load due to appliances that are always on, such as a refrigerator.

The third example is studying {\em daily load profile} of each customer. This algorithm is for extracting daily consumption patterns that occur regardless of outdoor weather temperatures (see Figure~\ref{fig:parx}). The left of the figure illustrates a fragment of the hourly consumption time series of a customer over a period of several days. Since smart meters report the total electricity consumption of a household, we can only observe the total consumption time series (the upper black curve). The goal of this algorithm is to determine how much load is due to temperature for each hour (\ie, heating and cooling), and how much load is due to daily activity independent of the temperature (the lower blue curve). Once this is determined, the algorithm fits a time series using auto-regression model, and computes the average temperature-independent consumption at each hour of the day, which is illustrated on the right of Figure~\ref{fig:parx} (the X-axis is the hour of the day, and the Y-axis is the average consumption). Since weekday and weekend activities may differ, it is useful to separately compute typical weekday and weekend profiles for each customer.

The last example is {\em anomaly detection} on the energy consumption of each customer. The anomaly detection can discover the consumption anomalies, such as an unusual high daily consumption, comparing with one's own consumption history. Figure \ref{fig:anomalydetect} shows an example of a customer's daily consumption at each hour. The actual consumption at 4 am is much higher than the predicted. Anomaly detection analytics many help to find the root causes of the unusual consumption, such as energy leakage, theft, or forgetting to turn off the stove after cooking, etc.

In following experimental studies, we will use the above four representative analytics algorithms to evaluate the proposed ICT-solution. The first three algorithms use the historical consumption data of each customer, which will be used to evaluate the batch analytics capability of the system; while the last one is a stream data mining algorithm that will be used to evaluate the real-time analytics capability.   
\afterpage{
\begin{figure*}[t!]
 \centering
 \subfigure[Variability analytics]{
\epsfig{file=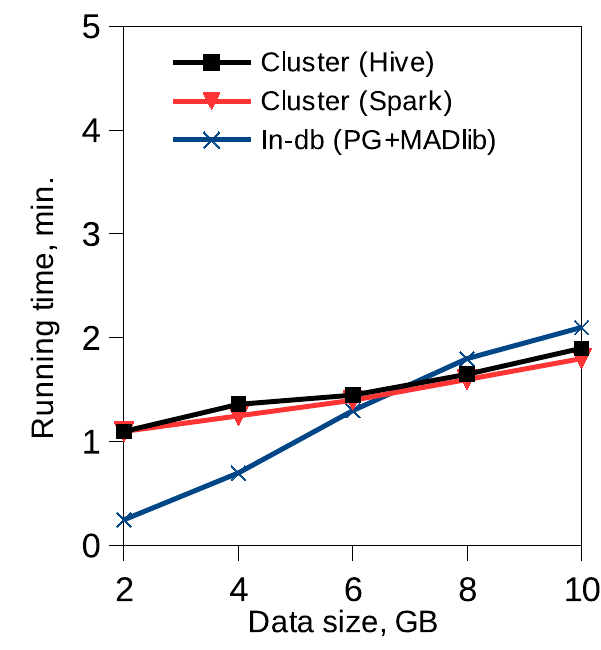, scale=0.7}
   \label{fig:histreal}
   }
 \subfigure[Thermal sensitivity analytics]{
\epsfig{file=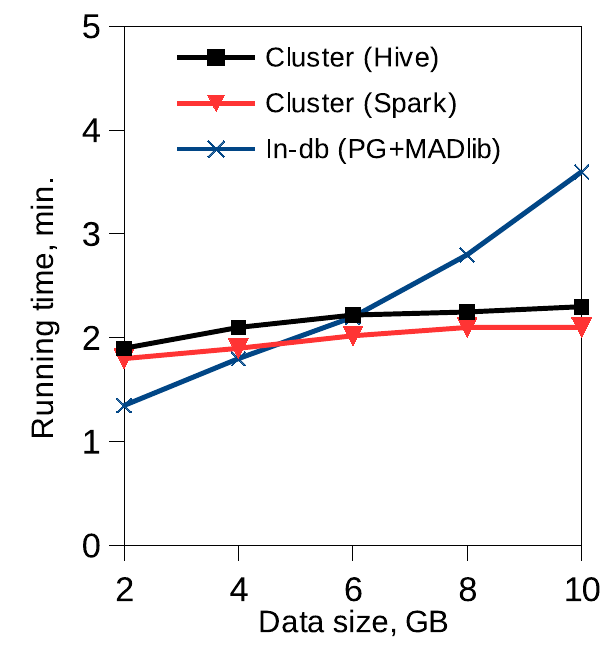, scale=0.7}
   \label{fig:threelinereal}
   }
 \subfigure[Daily load profile analytics]{
\epsfig{file=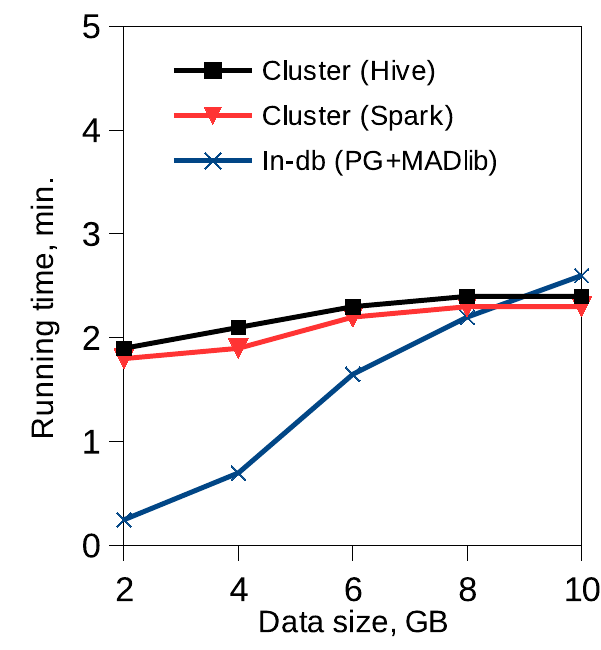, scale=0.7}
   \label{fig:parxreal}
   }
 \caption{Batch analytics performance using real-world data sets}
  \label{fig:batchreal}
% \vspace{-10pt}
\end{figure*}
\begin{figure*}[t!]
 \centering
  \subfigure[Variability analytics]{
\epsfig{file=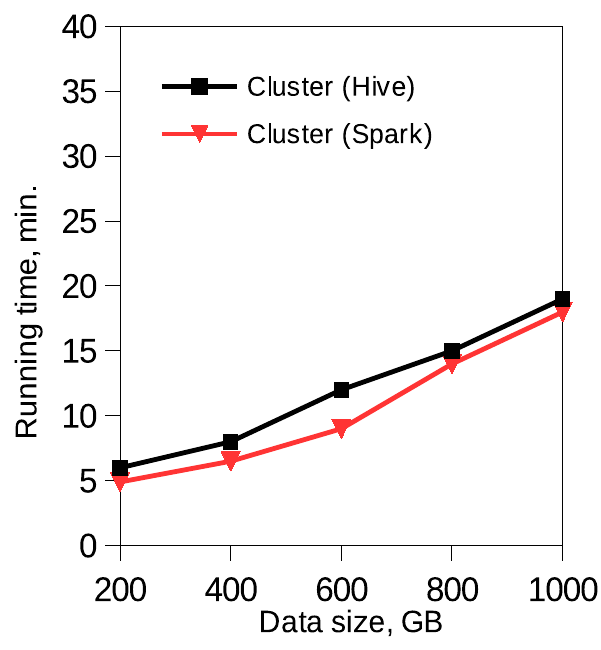, scale=0.7}
   \label{fig:histsyn}
   }
 \subfigure[Thermal sensitivity analytics]{
\epsfig{file=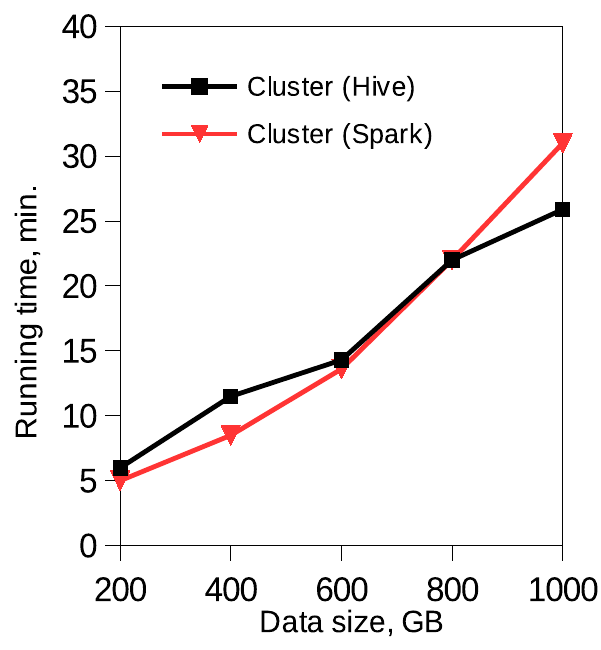, scale=0.7}
   \label{fig:threelinesyn}
   }
 \subfigure[Daily load profile analytics]{
\epsfig{file=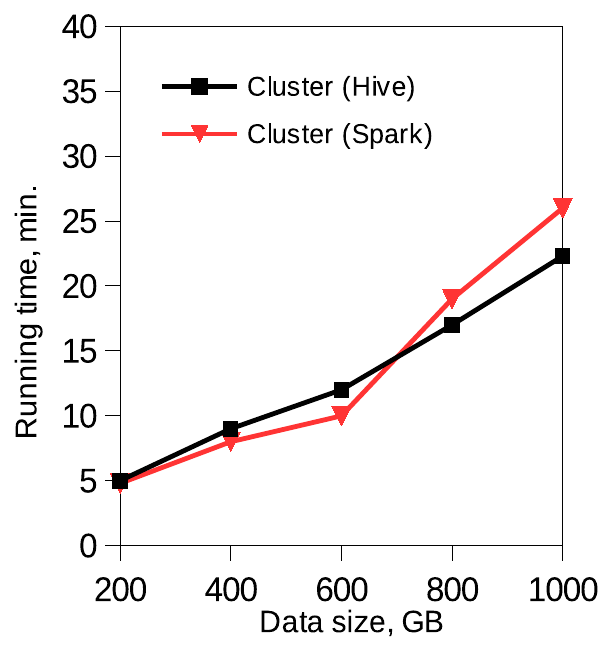, scale=0.7}
   \label{fig:parxsyn}
   }
 \caption{Batch analytics performance using synthetic data sets}
  \label{fig:batchsyn}
\end{figure*}
}

\subsection{Experimental Settings}
\label{sec:smartmeterdataanalytics}
In the following, we outline the experimental settings. The analytics layer with PostgreSQL and MADlib is installed on a single server while the data processing layer with Spark and Hive is installed in a cluster.  The analytics server is configured with an Intel(R) Core(TM) i7-4770 processor (3.40GHz, 4 Cores, hyper-threading is enabled, two hyper-threads per core), 16GB RAM, and a Seagate Hard driver (1TB, 6 GB/s, 32 MB Cache and 7200 RPM), running Ubuntu 12.04 LTS with 64bit Linux 3.11.0 kernel. PostgreSQL 9.1 has the settings ``shared buffers=3072MB, temp buffers= 256MB, work mem=1024MB, checkpoint segments =64'' and default values for other configuration parameters. The cluster consists of one administration node and 16 worker nodes. The administration node is the master node of Hadoop and HDFS, and clients submit jobs there. All the nodes have the same configuration: dual-socket Intel(R) Xeon(R) CPU E5-2620 (2.10GHz, 6 cores per socket, and two hyper-threads per core), 60GB RAM, running 64bit Linux with kernel version 2.6.32. The nodes are connected via gigabit Ethernet, and a working directory is NFS-mounted on all the nodes.

Real-world and synthetic data sets are both used for the evaluation. Since we only have 10GB real-world residential electricity consumption data sets, for more data we generate the synthetic data sets seeded by the real-world data. One gigabyte data set contains roughly 27,300 hourly granular time-series with the length of two years. The size of data tested in the cluster environment is scaled up to one terabyte, corresponding to over twenty million time series. 

In the following, we will first evaluate batch analytics using the technologies Hive and Spark (used in the processing layer), and PostgreSQL/MADlib (used in the analytics layer). Then, we will evaluate real-time analytics using the same technologies.

\subsection{Batch Analytics}
\label{sec:batchanalytics}
We evaluate cluster and in-database based batch analytics on BigETL (with Hive and Spark as the underlying data processing systems), and on PostgreSQL/MADlib using the first three illustrative examples, respectively. The implementations are that 1) {\em variability analytics:} we implement our histogram program in Java; 2) {\em thermal sensitivity analytics:} we use the functions from Apache common math library \citep{apachemath} to implement the three linear regression lines, and adjust the three piece-wise lines to connect them together; and 3) {\em daily load profile analytics:} the program is implemented based on the PARX model (see Section \ref{sec:indbanalysis}), which also uses the multiple linear regression function from Apache common math library. The implementations for the in-database analytics are the PostgreSQL stored procedures using MADlib analytics functions, including histogram and linear regression. 
 
We first use the real-world data set to evaluate batch analytics performance on the three platforms. We measure the total running time by scaling the data size from 2 to 10GB. In fact, this experiment is unfair in the sense that we test PostgreSQL with MADlib on a server (with the maximum parallelism level of eight hyper-threads), but test Spark and Hive on a cluster. To be more comparable, we make use of eight database connections in PostgreSQL to execute analytics queries in parallel, each of which is given the same number of time-series as the input. In the cluster, we make use of eight slave nodes, each of which runs a single task. Figure~\ref{fig:batchreal} shows the experimental results, indicating that the in-database analytics has better performance when handling relatively small-sized data while the cluster-based approach (both Hive and Spark) outperforms it for bigger sized data. The breaking points vary with the algorithms. This also verifies that the cluster-based approach is a better option for big data analytics in terms of the performance. In this experiment, we could observe that the running times used change insignificantly for both Hive and Spark (with flat lines) since the workload is too low for the cluster-based analytics. 

We now use big synthetic data to evaluate Hive and Spark, scaling from 200 to 1,000GB. All the 16 slave nodes are used in this experiment. Figure \ref{fig:batchsyn} shows the results, indicating that Spark has better performance in variability analytics, but Hive is better in thermal sensitivity analytics and daily load profile analytics after 700GB approximately. We found that this was due to the memory spilling occurring in Spark. Spark is an in-memory based distributed computing framework. If data objects (RDDs) cannot fit into the size of main memory, some of the data objects will be spilled over to hard drivers, which greatly deteriorates in performance. We have also observed that no data spilling happened in running the variability analytics algorithm. The reason is that the variability analytics only uses the consumption time series, while the other two algorithms also use the weather temperature time series, meaning that the required memory size is much bigger. 

If the performances of the algorithms are considered, the variability analytics shows to be better than the other two algorithms when using both types of data. The thermal sensitivity analytics algorithm needs more time than the daily load profile analytics algorithm. This can be explained by the following. Variability analytics uses the histogram which is simple and efficient; thermal sensitivity analytics requires to run the regression function three times for the three linear regression lines, while daily load profile analytics runs regression function only once.

\subsection{Real-time Analytics}
We now evaluate real-time analytics capability using online anomaly consumption detection. 

We use unique variate Gaussian distribution as our anomaly detection  model, which is defined in the following. Suppose that we have a training data set, $X=\{x_1, x_2, ..., x_n\}$,  whose data points obey normal distribution with the mean $\mu$ and the variance $\delta^2$. The detection function is defined as
\begin{equation}
p(x; \mu, \delta)=\frac{1}{\delta \sqrt{2\pi}}e^{-\frac{(x-\mu)^2}{2\delta^2}}
\end{equation}
where $\mu=\frac{1}{n} \sum^n_{i=1} x_i$ and $\delta^2=\frac{1}{n}\sum_{i=1}^{n}(x_i-\mu)^2$. If we have a new data point, $x$, we use this function to compute the probability density. If the probability is less than a use-defined threshold, \ie, $p(x)<\epsilon$,  it is classified as an anomaly. In our example, a data point is the Euclidean distance between the actual and predict hourly consumption of a day,  \ie,  $x_i = ||Y_i - \hat{Y}_i||$ where $Y_i$ is the $i$th day's actual hourly consumption defined as $Y_i=<y_0, y_1, ..., y_{23}>$, and $\hat{Y}_i$ is the predicted hourly consumption defined as $\hat{Y}_i=<\hat{y}_0,\hat{y}_1, ..., \hat{y}_{23}>$. We use the PARX model to compute the predicted hourly consumption of each day.

In our experiment, we simulate receiving hourly consumption of each day in a real-time fashion, \ie, the 24-hour consumption values are fed into Hive or Spark Streaming directly from Hadoop distributed file system (HDFS), while for PostgreSQL/MADlib, the values are fed from a table in PostgreSQL. Strictly speaking, only Spark Streaming is originally designed for processing the real-time data stream, e.g., connects to external streaming data sources, such as smart meters or sensors, and processes it. But, for comparison purpose, we also test Hive and MADlib by the simulation of reading data from their local data storage, \ie, tables in PostgreSQL or Hive. In our evaluation, we use half-year's data as the training data sets to compute the anomaly detection model, and we use the computed model throughout the subsequent detection process.     
\begin{figure}[htp]
\centering
\includegraphics[width=0.42\textwidth]{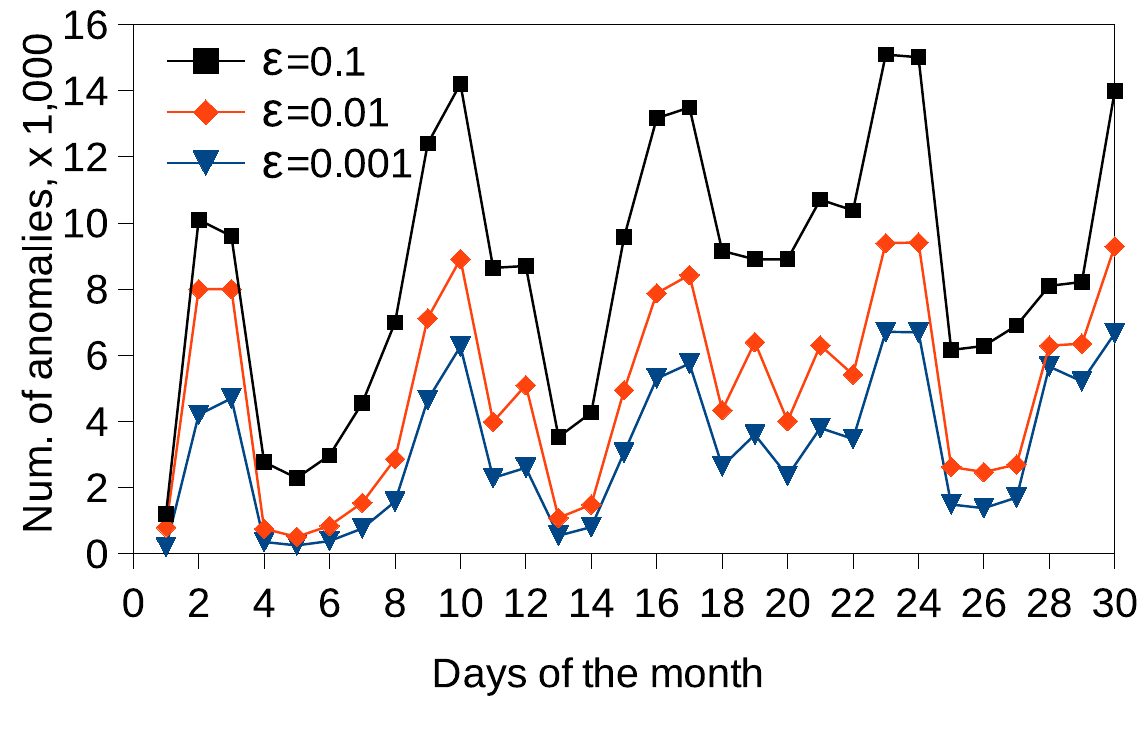}
\vspace{-15pt}
\caption{Anomaly detection results }
\label{fig:anomalyresults}
\end{figure}

\begin{figure*}[t!]
    \centering
    \begin{minipage}{.5\textwidth}
        \centering
        \includegraphics[width=0.8\linewidth]{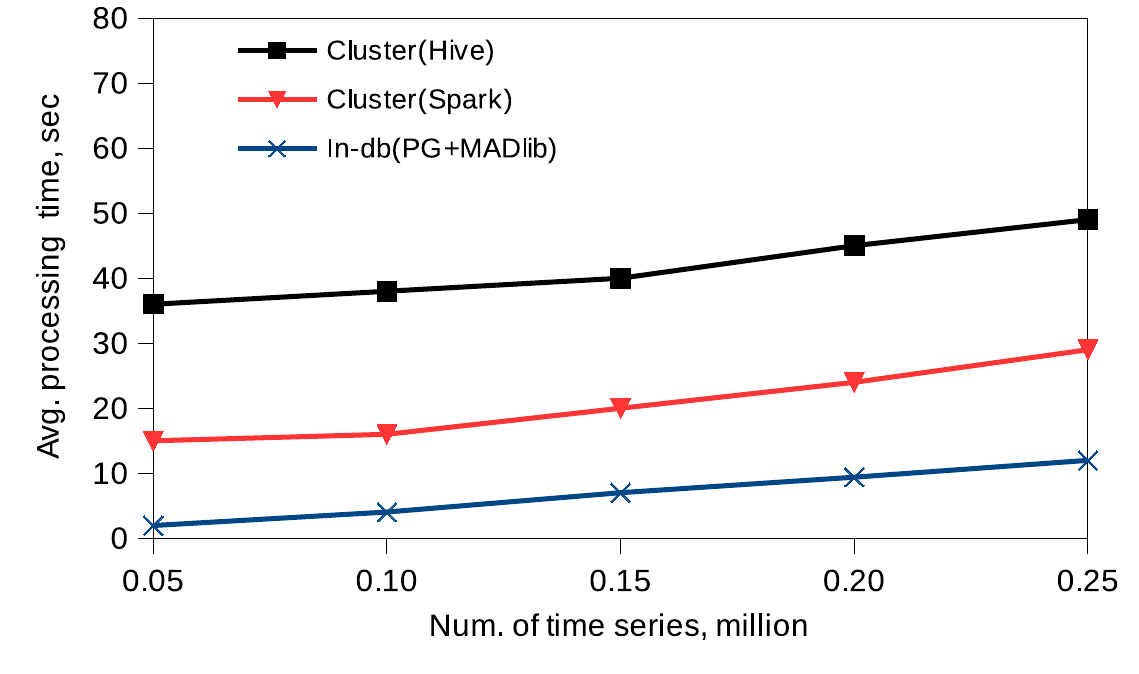}
        \caption{Real-time anomaly detection using real-world data sets}
        \label{fig:selfanomalyreal}
    \end{minipage}%
    \begin{minipage}{0.5\textwidth}
        \centering
        \includegraphics[width=0.8\linewidth]{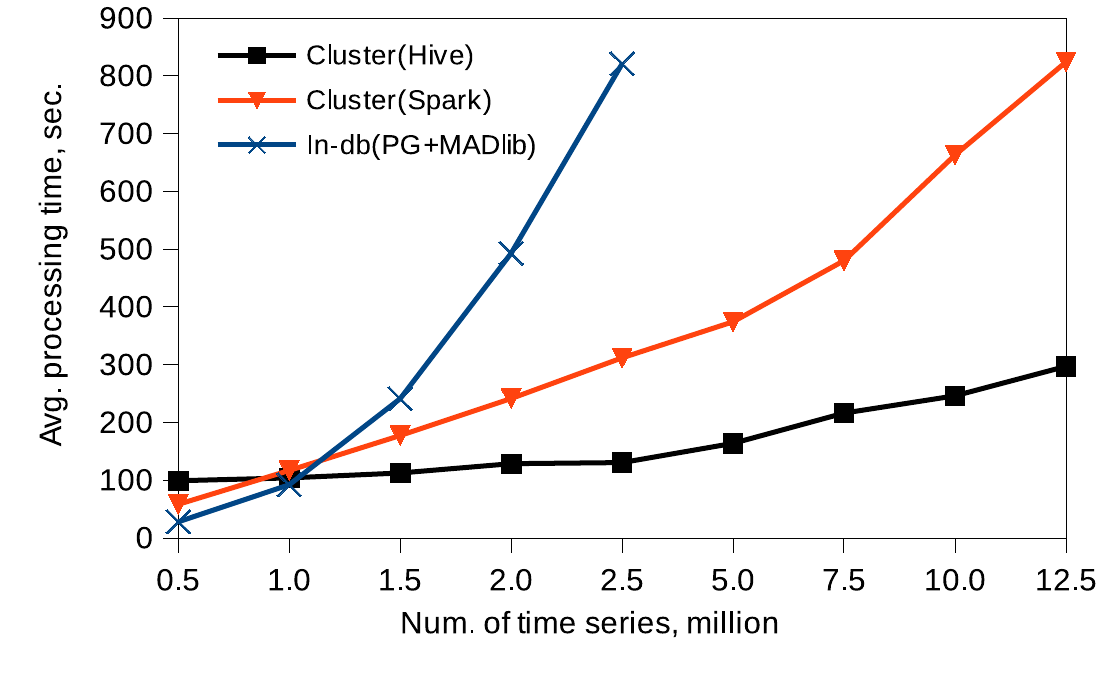}
        \caption{Real-time anomaly detection using synthetic data sets}
        \label{fig:selfanomalysyn}
    \end{minipage}
\end{figure*}
Here, we first use the real-world data sets again. We use one month's data (June 2012) as the example to test our algorithm. Figure~\ref{fig:anomalyresults} shows the detected anomalies when the threshold value, $\epsilon$, is set to, $0.1$, $0.01$ and $0.001$, respectively. As shown, the number of anomalies fluctuates in each day, and we could observe that the 2-3th, 9-10th, 16-17th, 23-24th and 30th days in this month have a larger number of anomalies than the other days. In fact, these days are the weekends when people might stay at home, and use more energy. The results also show that the number of anomalies is different when the threshold values are changed. The number gets closer when the threshold value decreases from $0.01$ to $0.001$. In this experiment, since we detect anomalies based on one's history consumption regardless of the weekends or weekdays, a relatively high consumption in weekends may be classified as an anomaly. This, however, can be improved by classifying the consumption values according to weekdays and weekends/holidays; or by comparing clustered customer groups or neighborhood. Besides, it would be favorable for customers to set their threshold values, e.g., through a mobile phone or web client, to decide when they could receive alerting messages. We leave these improvements to our future work.

We now measure the performance. First, we use the real-world data sets by scaling the number of time series from 5,000 to 250,000 and measure the average time of processing the data of each day. Figure \ref{fig:selfanomalyreal} shows the results of using the cluster based (Hive and Spark Streaming), and the in-database based (PostgreSQL/MADlib) approaches. As illustrated, the in-database based approach outperforms the cluster based, and Spark Streaming shows a better performance than Hive in the cluster based approach, due to its use of in-memory-based technology. Obviously, they all are under workload at these scales of the data set. 
 
We now use the synthetic data set, and scale the number of time series from 0.5 to 12.5 million (up to 457GB). Figure \ref{fig:selfanomalysyn} shows the results, where Hive is the best after  1.0 million time series, and the in-database approach becomes less efficient after 1.2 million. It is not surprising that the in-database approach becomes slower for large data sets since it runs on a single server which has a limited scalability. For the cluster based approach, when a relatively small number of time series is given as the input, the performance difference is likely due to overheads associated with task distribution; by default, Hive launches a separate Java Virtual Machine (JVM) for each task whereas Spark reuses task executors more intelligently. In fact, even running a ``SELECT *" Hive query from a one-row table took nearly 30 seconds in our cluster. When a bigger number of time series is given, \ie, $>1.2million$, Spark performs substantially worse due to the overhead of accessing data from the past three days (recall that we use the order $p=3$ in PARX model for the prediction, see Section \ref{sec:indbanalysis}). Spark Streaming checkpoints the sliding window of the past three days to HDFS whenever new data are added, which lowers the performance.  

\subsection{Discussion}
We have evaluated the key technologies that constitute the proposed ICT-solution. The results reveal that the in-database analytics using PostgreSQL/MADlib is more suitable for relatively small-sized data sets, and can provide highly responsive interactive analytics, due to the support of high-performance OLTP operations and indexing in a database management system. For big data analytics, it is favorable to go for a cluster-based approach, e.g., using Hive or Spark, to obtain high scalability. In the cluster based analytics, Spark shows a better performance than Hive in general on the condition that the in-memory data is not spilled over to local hard-driver on a node. Therefore, it is beneficial to provide sufficient memory to ensure the performance in Spark. The proposed ICT-solution employs a hybrid architecture in order to get the best of each of the systems, \ie, high-performance analytics queries of {\em PostgreSQL}, big data capability of {\em Hive} and {\em Spark}, and real-time ability of Spark Streaming. By the hybrid architecture, the ICT platform can achieve both batch and (near) real-time analytics for smart meter data analytics. 
 
Based on our studies, we recommend using PostgreSQL/MADlib to manage the latest smart meter data (\eg, of the past two years), social-economic data, and statistic data. Smart meter data sets are typically big, but rarely updated; socio-economic data sets such as the information of customers are typically much smaller, but may frequently be updated; Statistic data sets are the result of analytics algorithms, which are also typically small. Furthermore, when data get older, they will usually lose the values. Thus, old data can be moved to the offline data warehouse system, Hive. However, when the data are needed, users still can run batch analytics queries directly on Hive, and transfer the results to the online data warehouse in PostgreSQL for the interactive analytics. For (near) real-time data analytics, Spark Streaming supports stream processing in nature, which reads stream data at a regular time interval, and uses stream operations to process the data, such as sliding window and stream join operations. In contrast, if users want to use PostgreSQL/MADlib or Hive, they have to implement their own data streaming programs, which require much more effort. 
 
The current focus of this platform is for smart grid data analytics, but it can easily be extended to support distribution network management. Smart grids such as low-voltage (LV) networks are characterized by variable demands, which requires integrating distributed renewable energy sources and energy storages for balancing energy supply and demand \citep{bennett2014,bennett2015}. The platform is suitable for energy storage scheduling management. For example, this platform can flexibly integrate different forecasting algorithms, e.g., ARIMA, PARX, ARIMAX \citep{bennett20142}, to predict day-/hour-ahead energy consumption profile. Then based on the load profile, the scheduling system can optimize supply-side energy flexibility using a dynamic scheduling strategy, \ie, charging batteries during low-demand periods while discharging during peak-demand periods. The application in smart grid distribution network management will be our future work.

\section{Conclusions and Future Work}
\label{sec:conclusionandfuturework}
With the widely implementation of smart meters, smart meters produce considerable volumes of data, presenting the opportunity for utilities to enhance customer service, lower cost and improve energy efficiency; and for customers to save energy and reduce the bills. Smart meter data analytics is a complex process that involves data ingestion, pre-processing, analytics, and visualization. In this paper, we proposed an ICT-solution to streamline smart meter data analytics. The proposed ICT solution employs a hybrid system architecture combining different technologies, including Spark, Hive, and PostgreSQL/MADlib, etc. The system architecture consists of three layers, including ingestion layer, processing layer, and analytics layer, each of which supports the extension for different data processing and analytics purposes. In particular, we introduced in-database analytics to achieve high performance and cluster-based big data analytics in our system. The ICT-solution can handle both (near) real-time and batch analytics for smart meter data. We have tested the effectiveness and efficiency of the ICT-solution comprehensively using real-world and synthetic data sets. The results have shown that the proposed solution can analyze batch and stream data effectively and efficiently.

In the future work, we will improve the system by adding new features, and more analytics algorithms developed in our research. We would like to extend our system to support other types of smart meter data, such as water, gas, and heat; and to extend our system for the support of smart grid distribution network management that we have discussed. Moreover, we will investigate the potential applications for utilities based on the results provided in this paper; and study how this system help utilities in their energy management.  

\section*{Acknowledgments} 
This research was supported by the CITIES project funded by Danish Innovation Fund (1035-00027B). 

\section*{References} 
\bibliographystyle{elsarticle-num}

\begin{thebibliography}{20}
\balance



\bibitem{msw10}
Greentech Media Research, The Soft Grid 2013-2020: Big data \& utility analytics for smart grid.  \url{http://www.greentechmedia.com/research/report/the-soft-grid-2013} as of  2016-05-11.


\bibitem{sdewe2015}
Liu X, Nielsen P S. Streamlining smart meter data analytics. In Proc. of the 10th Conference on Sustainable Development of Energy, Water and Environment Systems, SDEWES2015.0558, 1-14, 2015.


\bibitem{bigetl}
BigETL -- A scalable data processing system. \url{http://github.org/xiufengliu/bigetl} as of 2016-05-11.

\bibitem{smas2015}
Liu X, Golab L, Ilyas IF. SMAS: a smart meter data analytics system. Proc. of ICDE 2015; 1476--1479.

\bibitem{madlib}
MADlib.  \url{madlib.net} as of 2016-05-11.


\bibitem{benchmark2015}
Liu X, Golab L,  Golab W, Ilyas IF. Benchmarking smart meter data analytics. Proc. of EDBT 2015; 385--396.

\bibitem{kdb}
KDB+ -- The time-series database for performance-critical environments.  \url{http://kx.com/} as of 2016-05-11.


\bibitem{farber2012}
F\'{a}rber F,  Cha SK, Primsch J, Bornh\"{o}vd C, Sigg S, Lehner W.  SAP HANA database: data management for modern business applications. ACM Sigmod Record 2012; 40(4):45--51.

\bibitem{Personal2014}
\new{Personal E, Guerrero JI, Garcia A, Pena M, Leon C. Key performance indicators: A useful tool to assess smart grid goals. Energy 2014; 76:976-988}

\bibitem{smartd}
Nezhad AJ,  Wijaya TK, Vasirani M,  Aberer K. SmartD: Smart meter data analytics dashboard.  Proc. of Future Energy Systems (ACM e-Energy) 2014; 213--214.


\bibitem{liu2014}
Liu Y, Hu S, Rabl T,  Liu W, Jacobsen HA, Wu K, Chen J.  DGFIndex for smart grid: enhancing Hive with a cost-effective multidimensional range index. PVLDB 2014; 1496--1507.

\bibitem{stewart2015}
Stewart RA, Willis R, Giurco D, Panuwatwanich K, Capati G. Web-based knowledge management system: linking smart metering to the future of urban water planning. Australian Planner 2010; 47(2):66--74.

\bibitem{khoi2015}
Nguyena KA, Stewart RA, Zhang H, Jones C. Intelligent autonomous system for residential water end use classification: Autoflow. Applied Soft Computing 2015; 31:118--131.



\bibitem{arlitt2015}
Arlitt M, Marwah M, Bellala G, Shah A, Healey J, Vandiver  B. IoTAbench: an Internet of Things analytics benchmark. Proc. of ICPE 2015.

\bibitem{keogh2003}
Keogh E, Kasetty S. On the need for time series data mining benchmarks: a survey and empirical demonstration. data miningand know. Disc. (DMKD) 2003; 7(4):349--371.


\bibitem{anil2013}
Anil C. Benchmarking of data mining techniques as applied to power system analysis. Master Thesis 2013; Uppsala University.

\bibitem{birt}
Birt BJ,  Newsham GR,  Beausoleil-Morrison I, Armstrong MM,  Saldanha N, Rowlands IH. Disaggregating categories of electrical energy end-use from whole-house hourly data. Energy and Buildings 2012; 50:93--102.


\bibitem{rvn10}
Rasanen T, Voukantsis D,  Niska H, Karatzas K,  Kolehmainen M. Data-based method for creating electricity use load profiles using large amount of customer-specific hourly measured electricity use data. Applied Energy 2010; 87(11):3538--3545.

\bibitem{zeng20141}
\new{Zeng C, Wu C, Zuo L, Zhang B, Hu X. Predicting energy consumption of multiproduct pipeline using artificial neural networks. Energy 2014; 66:791--8.}

\bibitem{acf12}
Abreu JM, Camara FP,  Ferrao P. Using pattern recognition to identify habitual behavior in residential electricity consumption. Energy and Buildings 2012; 49:479-487.

\bibitem{Gebru}
\new{Arghira N, Hawarah L, Ploix S, Jacomino M. Prediction of appliances energy use in smart homes. Energy 2012; 48(1):128--134.}

%\bibitem{Gebru}
%Albert A, Gebru T, Ku J, Kwac J,  Leskovec J, Rajagopal R. Drivers of Variability in Energy Consumption. Proc. of ECML-PKDD DARE Workshop on Energy Analytics 2013.


\bibitem{ar13}
Albert A, Rajagopal R.  Smart meter driven segmentation: what your consumption says about you. IEEE Transactions on Power Systems 2013; 4(28).

\bibitem{omid}
Ardakanian O, Koochakzadeh N, Singh RP, Golab L,  Keshav S. Computing electricity consumption profiles from household smart meter data. Proc. of EnDM Workshop on Energy Data Management 2014; 140--147.

\bibitem{beckel2014}
Beckel C, Sadamori L, Staake T,  Santini, S.  Revealing household characteristics from smart meter data. Energy 2014, 78:397--410.

\bibitem{chicco}
Chicco G, Napoli R, Piglione F. Comparisons among clustering techniques for electricity customer classification. IEEE Trans. on Power Systems 2006; 21(2):933-940.


\bibitem{espinoza}
Espinoza M,  Joye C,  Belmans R, DeMoor B. Short-term load forecasting, profile identification, and customer segmentation: a methodology based on periodic time series. IEEE Trans. on Power Systems 2005; 20(3):1622--1630.

\bibitem{frv05}
Figueiredo V, Rodrigues F, Vale Z,  Gouveia  J. An electric energy consumer characterization framework based on data mining techniques. IEEE Trans. on Power Systems 2005; 20(2):596--602.

\bibitem{ghe11}
Ghofrani M, Hassanzadeh M, Etezadi-Amoli M,  Fadali M. Smart meter based short-term load forecasting for residential customers. North American Power Symposium (NAPS) 2011.

\bibitem{smith}
Smith BA, Wong J, Rajagopal R. A simple way to use interval data to segment residential customers for energy efficiency and demand response program targeting. ACEEE Summer Study on Energy Efficiency in Buildings 2012.


\bibitem{tsekouras}
Tsekouras G, Hatziargyriou N, Dialynas E. Two-stage pattern recognition of load curves for classification of electricity customers. IEEE Trans. Power Systems 2007; 22(3):1120--1128.

\bibitem{albert_bigdata}
Albert A, Rajagopal R. Building dynamic thermal profiles of energy consumption for individuals and neighborhoods. Proc. of IEEE Big Data 2013; 723--728.

\bibitem{etlmr2011}
Liu X, Thomsen C,  Pedersen TB. ETLMR: a highly scalable etl framework based on MapReduce. Proc. of DaWaK 2011; 96--111.

\bibitem{tldks2011}
Liu X, Thomsen C,  Pedersen TB. ETLMR: A Highly Scalable Dimensional ETL Framework Based on MapReduce. TLDKS III 2013; 8:1-31.

\bibitem{pvldb2011}
Liu X, Thomsen C,  Pedersen TB. MapReduce-based Dimensional ETL Made Easy (demo). PVLDB 2012; 5(12): 1882-1885.

\bibitem{cloudetl2014}
Liu X, Thomsen C,  Pedersen TB. CloudETL: scalable dimensional ETL for Hive. Proc. of IDEAS 2014;  195-206.




\bibitem{sparkstreaming}
Zaharia M, Das T, Li H, Shenker S, and  Stoica I. Discretized streams: an efficient and fault-tolerant model for stream processing on large clusters.  Proc. of HotCloud 2012; 10--10.

\bibitem{bigsql}
BigSQL. \url{http://www.bigsql.org} as of 2016-05-11.


\bibitem{hive2010}
Thusoo A, Sarma JS, Jain N, Shao Z, Chakka P, Zhang N, Murthy R. Hive-a petabyte scale data warehouse using Hadoop. Proc. of ICDE 2010; 996--1005.


\bibitem{apachemath}
Apache common math library. \url{https://commons.apache.org/proper/commons-math/} as of 2016-05-11.

\bibitem{bennett2014}
Bennett CJ, Stewart RA, and Lu JW.  Forecasting low voltage distribution network demand profiles using a pattern recognition based expert system. Energy 2014; 67:200-212.

\bibitem{bennett2015}
Bennett C J, Stewart RA, and Lu JW. Development of a three-phase battery energy storage scheduling and operation system for low voltage distribution networks. Applied Energy 2015; 146:122--134.

\bibitem{bennett20142}
Bennett CJ, Stewart RA, and Lu JW.  Autoregressive with exogenous variables and neural network short-term load forecast models for residential low voltage distribution networks. Energies 2014; 7(5):2938-2960.



%\bibitem{spark}
%Apache Spark. \url{spark.apache.org} as of 2016-05-11.


%\bibitem{madlibpaper}
%Hellerstein JM, Re C, Schoppmann F, Wang DZ, Fratkin E, Gorajek A, Kumar A. The MADlib Analytics Library: or MAD Skills, the SQL. PVLDB 2012; 5(12):1700--1711.

%\bibitem{zaharia2010}
%Zaharia M, Chowdhury M, Franklin MJ, Shenker S, Stoica I. Spark: Cluster Computing with Working Sets. Proc. of HotCloud 2010.


\end{thebibliography}
%\begin{thebibliography}{20}

\appendix
\section{The Validation of PARX}
\subsection{Coefficient validity}
\label{apx:parx}

We randomly take a consumption time series from the real-world data set to evaluate PARX model, and the output is given in Table~\ref{tbl:coef}. According to the results, \ie, p-test, the coefficient estimates of the most recent three-day's consumption values ($p=3$) at the day $d$ have shown a good significance, which is same to the temperature coefficient estimates. The number of ``*" shows the significance level.
\begin{table}[htp]
\centering
{\scriptsize
\caption{The validity of the coefficients of the PARX model}
\label{tbl:coef}
\begin{tabular}{p{1cm}p{1cm}p{1cm}p{1cm}p{1.1cm}p{1cm}}
\hline
Explanatory variable  & Coefficient estimate  & Std. error  & t-value & Two-tailed p-test & Significance \\ \hline
Intercept             & 0.504        & 0.0729      & 6.92    & 1.33e-11          & ***          \\
$y_{d-1}$                 & 0.316        & 0.0406      & 7.79    & 3.58e-14          & ***          \\
$y_{d-2}$                  & 0.108        &   0.0387    & 3.45     &0.001            & **            \\
$y_{d-3}$                  & 0.133        & 0.0422       & 2.56    &  0.0107          & *           \\
$XT1$                   & 0.194        & 0.0189      & 10.24   & 1.34e-22          & ***          \\
$XT2$                   & -0.029       & 0.0085      & -3.38   & 0.001             & **           \\
$XT3$                   & 0.052        & 0.0193      & 2.68    & 0.008             & **           \\ \hline
\multicolumn{6}{l}{0 '***', 0.001 '**', 0.01 '*', 0.05 '.', 0.1 ''  Adjusted $R^2$: 0.6341, n=534}
\end{tabular}
}
%\vspace{10pt}
\end{table}

\subsection{Accuracy validation}
\label{apx:parx}
We randomly select 10\% consumption time series (2,730) from the real-world data set to evaluate the predictive accuracy. For each time series, we use a quarter of the readings as the training data to create the model (\ie, six months), and the rest as the testing data. During the test, the model is refreshed iteratively by days, and the training set is expanded by adding the readings from the testing set prior to the day of testing. For better assessment, we compare PARX with the following three algorithms: 1) {\em Averaging:} we use the averaging value at a particular hour in the training data to predict the reading of the hour; 2) {\em 3-Line}: we use the three piece-wise linear regression line algorithm to predict the consumption with a given weather temperature (see  Figure~\ref{fig:3line}); and 3) {\em Convergent Vector}: this algorithm is proposed by \citep{espinoza}, which is similar to PARX taking weather temperature into account. For all the methods, we compute the root-mean-square error (RMSE)  between the actual and predicted values, which is defined as follows:
\begin{equation}
	RMSE=\sqrt{\frac{1}{n}\sum_{i=1}^{n}(\hat{Y}_i - Y_i)^2}
\end{equation}
where $\hat{Y}_i$ is the predicted value, $Y_i$ is the actual value, and $n$ is the size of testing data.
We compare the RMSE values of each time-series for the four prediction methods, and get the following findings:

\begin{itemize}
\item PARX outperforms Averaging for 2,586 time series, 3-Line for 2,612 time series, and Convergent Vector for 2,460 time series.

\item Table~\ref{tbl:rmse} summarizes the mean value of RMSE over all the 2,730 time series for each algorithm. PARX is 13.3\% lower than the Averaging, 21.7\% lower than 3-Line, and 6.0\% lower than Convergent vector.
\begin{table}[h]
\centering
{\scriptsize
\caption{The average RMSE value over 2,730 time series}
\label{tbl:rmse}
\begin{tabular}{p{1.5cm}p{1.5cm}p{1.5cm}p{2cm}}
\hline
PARX  & Averaging  & 3-Line & Convergent vector  \\ \hline
0.72  &0.83        & 0.92      & 0.76                    \\ \hline
\end{tabular}
}
\vspace{-10pt}
\end{table}

\end{itemize}
Therefore, according to the above results, PARX can outperform the other representative models, and has the lowest predictive errors on average. This confirms that the necessary of incorporating the seasonality of history consumption and  temperature dependence into a prediction model. 
%\vspace{-15pt}
\section{The web-based user interface}
\label{apx:screenshot}
\begin{figure*}[htp]
\vspace{-15pt}
\centering
\includegraphics[width=0.82\textwidth]{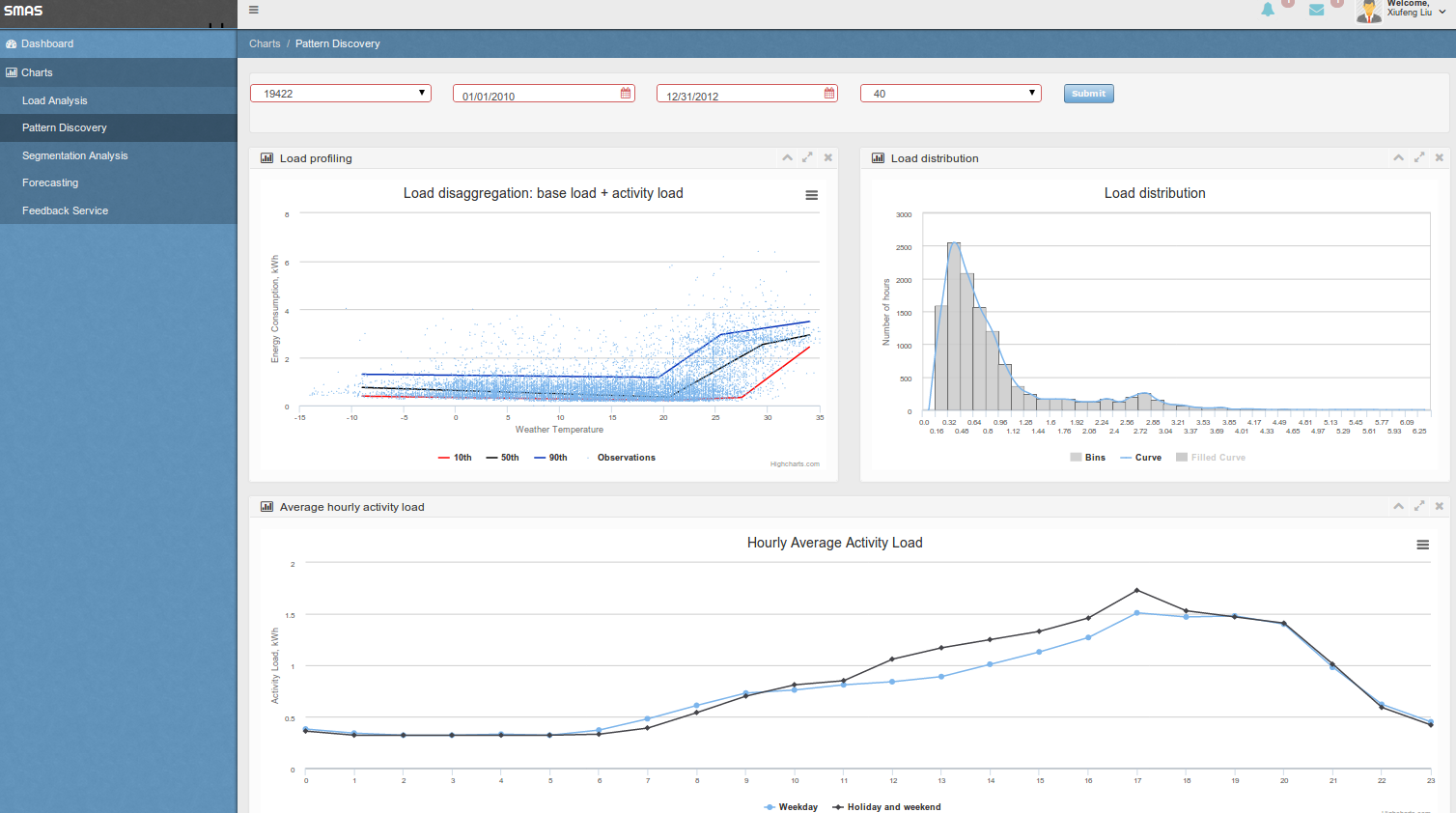}
\vspace{-10pt}
\caption{A screenshot of consumption pattern discovery}
\label{fig:patterndiscovery}
\vspace{-15pt}
\end{figure*}
\begin{figure*}[htp]
\vspace{-25pt}
\centering
\includegraphics[width=0.82\textwidth]{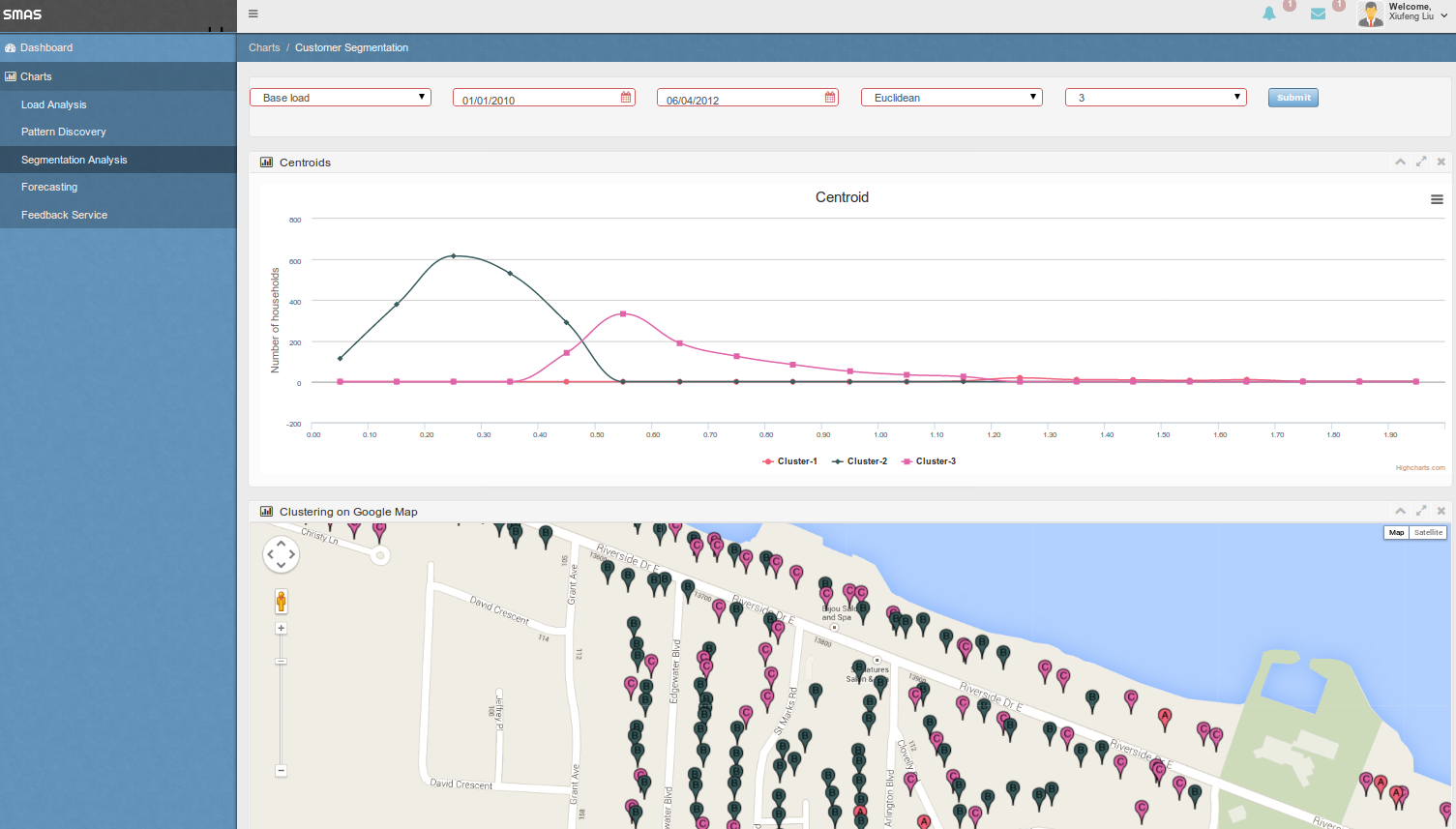}
\vspace{-10pt}
\caption{A screenshot of segmentation}
\label{fig:segmentation}
\vspace{-15pt}
\end{figure*}
\begin{figure*}[htp]
\vspace{-30pt}
\centering
\includegraphics[width=0.82\textwidth]{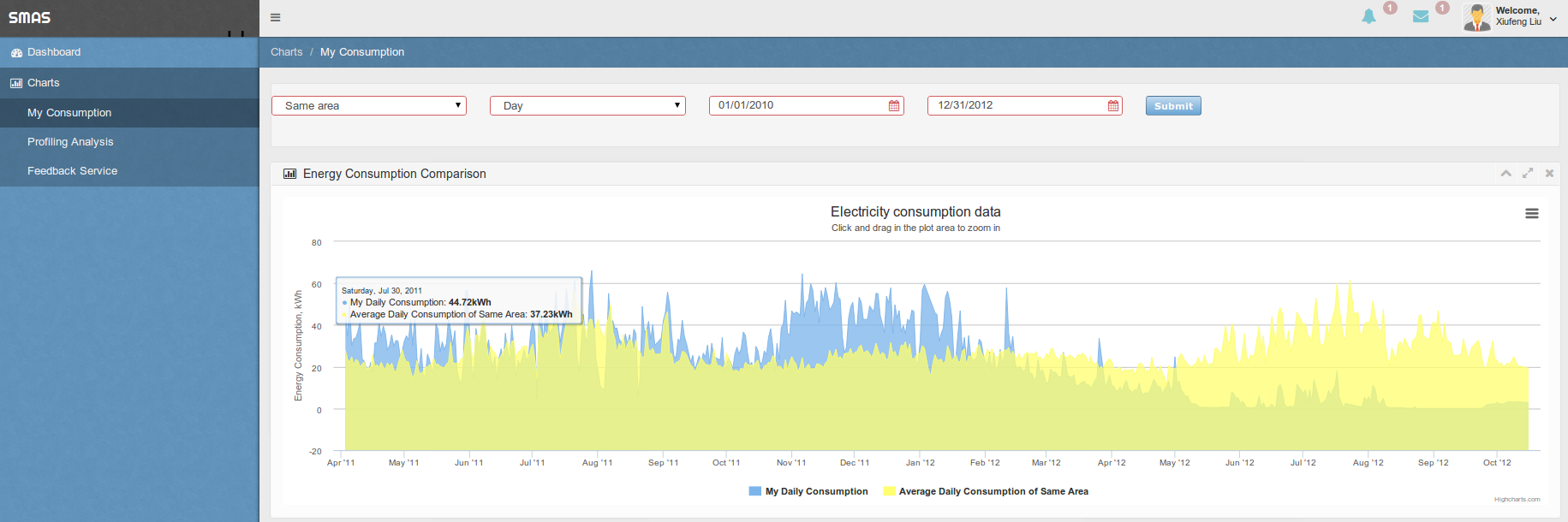}
\vspace{-10pt}
\caption{A screenshot of consumption analysis}
\label{fig:myconsumption}
\vspace{-40pt}
\end{figure*}
\end{document}